\documentclass[fleqn,usenatbib]{mnras}

\usepackage[T1]{fontenc}
\usepackage{color}

\DeclareRobustCommand{\VAN}[3]{#2}
\let\VANthebibliography\thebibliography
\def\thebibliography{\DeclareRobustCommand{\VAN}[3]{##3}\VANthebibliography}
\def\code#1{\texttt{#1}}

\usepackage{graphicx}	
\usepackage{amsmath}	
\usepackage{graphics,subfig,epsfig,amssymb,color,times}
\usepackage{hyperref}
\usepackage{comment}
\usepackage{newtxtext,newtxmath}
\usepackage{caption}
\usepackage{orcidlink}

\title[X-ray polarimetry of NGC~4151]{Uncovering the geometry of the hot X-ray corona in the Seyfert galaxy NGC~4151 with IXPE}

\author[IXPE Collaboration]{
V. E. Gianolli,$^{1,2}$\thanks{E-mail: vittoria.gianolli@univ-grenoble-alpes.fr}\orcidlink{0000-0002-9719-8740} D. E. Kim,$^{3,4,5}$\thanks{E-mail: dawoon.kim@inaf.it}\orcidlink{0000-0001-5717-3736} S. Bianchi,$^{2}$\orcidlink{0000-0002-4622-4240} B. Agís-González,$^{6}$\orcidlink{0000-0001-7702-8931} G. Madejski,$^{7}$ F. Marin,$^{8}$\orcidlink{0000-0003-4952-0835} \newauthor \; A. Marinucci,$^{9}$\orcidlink{0000-0002-2055-4946} G. Matt,$^{2}$\orcidlink{0000-0002-2152-0916} R. Middei,$^{3,10}$\orcidlink{0000-0001-9815-9092} P-O. Petrucci,$^{1}$\orcidlink{0000-0001-6061-3480}
% ...
\\
\\
     Affiliations are shown at the end of the paper 
}

\date{Accepted XXX. Received YYY; in original form ZZZ}

\pubyear{2023}

\begin{document}
\label{firstpage}
\pagerange{\pageref{firstpage}--\pageref{lastpage}}
\maketitle

\begin{abstract}
We present an X-ray spectro-polarimetric analysis of the bright Seyfert galaxy NGC~4151. The source has been observed with the {\it Imaging X-ray Polarimetry Explorer} ({\it IXPE}) for 700~ks, complemented with simultaneous {\it XMM-Newton} (50 ks) and {\it NuSTAR} (100 ks) pointings. A polarization degree $\Pi=4.9\pm1.1$\% and angle $\Psi=86\degr\pm7\degr$ east of north (68\% confidence level) are measured in the 2--8 keV energy range. 
The spectro-polarimetric analysis shows that the polarization could be entirely due to reflection. Given the low reflection flux in the {\it IXPE} band, this requires however a reflection with a very large ($>38$\%) polarization degree. Assuming more reasonable values, a polarization degree of the hot corona ranging from $\sim4$ to $\sim8$\% is found. 
The observed polarization degree excludes a `spherical' lamppost geometry for the corona, suggesting instead a \textit{slab}-like geometry, possibly a \textit{wedge}, as determined via Monte Carlo simulations. This is further confirmed by the X-ray polarization angle, which coincides with the direction of the extended radio emission in this source, supposed to match the disc axis. 
NGC~4151 is the first AGN with an X-ray polarization measure for the corona, illustrating the capabilities of X-ray polarimetry and {\it IXPE} in unveiling its geometry.

\end{abstract}




\section{Introduction}\label{introduction}

The common paradigm for Active Galactic Nuclei (AGN) \citep{antonucci93} postulates the presence of a corona of hot electrons ($kT_{\rm e} \simeq$~10--100 keV), responsible for the primary continuum in the hard X-rays through Inverse Comptonization of UV photons \citep{sunyaev80,zdziarski00}. Despite widespread acceptance of this process, the source of energy for the plasma and the conditions leading to its formation remain open questions. The geometry of this region further contributes to these debates, ranging from a \textit{slab}-corona model \citep{haardt91,haardt93,merloni03}, in which the energy dissipation and electron heating occur over a large volume, to a compact source located on the accretion disc axis \citep[lamppost geometry,][]{martocchia96, fabian17} and whose possible origin could be an aborted jet \citep[see][]{henri97,ghisellini04}.
Compton scattering will produce a polarization signal which is strongly sensitive to the geometry of the scattering material. 
Although spectroscopy and timing alone were not been able to distinguish between geometrical models so far, their application alongside polarization can aid in determining the characteristics of the corona such as the Thomson optical depth $\tau$ and the electron temperature $kT_{\rm e}$  \citep{shapiro76}.
X-ray polarimetry is thus a powerful tool that can be used to bring new insights on the innermost regions of AGN.
In particular, from `spherical' lamppost coronae, a polarization degree of one percent or so is expected, while larger values are anticipated when the scattering medium is distributed as a \textit{slab} over the accretion disc \citep{poutanen96,tamborra18,ursini22}. 

The Imaging X-ray Polarimetry Explorer \citep[{\it IXPE}, ][]{weisskopf22}, launched on December 9, 2021, is a NASA/ASI mission and the first X-ray imaging polarimeter in orbit after 40 years. Thanks to three telescopes with polarization-sensitive imaging detectors \citep[gas-pixel detector,][]{costa01} effective in the 2–8 keV energy band, X-ray polarimetric studies on AGN are being carried out for the first time. So far a total of four radio-quiet AGN (i.e., MCG-05-23-16, the Circinus galaxy, NGC~4151 and IC 4329A) have been observed by {\it IXPE}. A polarization degree $\Pi<4.7$\%  was derived for MCG-05-23-16, in agreement with expectations from a lamppost `spherical' geometry of the corona, or a \textit{slab} geometry if the inclination angle of the system is less than 50\degr\ \citep{marinucci22}. On the other hand, the Circinus galaxy shows a very high $\Pi \sim$ 28 $\pm$ 7 \% with a polarization angle perpendicular to the radio jet at about $\psi_{\rm X}\sim$ 18 $\pm$ 5\degr\ \citep{ursini23}. However, this source is Compton-thick with no direct view of the corona, so all the polarization is ascribed to reflection from an equatorial torus, as expected from the standard Unification Model.

NGC~4151 is one of the brightest Seyfert galaxies in the local universe. It has been classified as a Changing Look AGN \citep{penston84, puccetti07,shapovalova08}, going from optical type 1.5 at high flux states (in which the source reaches up to $F_{\rm{0.5-10 \; keV}}$ $\sim$ 2.8$\times$ 10$^{-10}$ erg s$^{-1}$ cm$^{-2}$) to optical type 1.8 at low fluxes states \citep[$F_{\rm{0.5-10 \; keV}}$ $\sim$ 8.7$\times$ 10$^{-11}$ erg s$^{-1}$ cm$^{-2}$; see][]{antonucci83,shapovalova12,Beuchert2017}. NGC~4151 has been intensively observed by all major X-ray satellites. It is characterised with significant spectral variability, and a complex absorption structure, both from neutral and ionised gas \citep[e.g.][]{Beuchert2017}. Below $\sim$2 keV, the soft X-ray emission is dominated by emission lines \citep[e.g.][]{Schurch2004}, likely arising from photoionizated gas in the narrow-line region, as commonly found in obscured AGN \citep{bianchi06,Guainazzi2007,Bianchi2019}. Previous studies found evidence for relativistic reflection off the accretion disc, suggesting a near-maximal spinning black hole \citep{cackett14, keck15, Beuchert2017}. Given a black hole (BH) mass of $4.57 \times 10^7 M_\odot$ \citep[from optical and UV reverberation,][]{bentz06}, the source has a relatively low Eddington ratio, 1\% \citep{keck15}.

In the following, we present the spectral and spectro-polarimetric analysis of the combined data from {\it IXPE}, {\it XMM-Newton} and {\it NuSTAR}, providing the most complete view to date of the inner accretion flow in NGC~4151. 
The paper is organized as follows. In Sect.~\ref{datareduction}, we describe the {\it IXPE}, {\it XMM-Newton} and {\it NuSTAR} observations and data reduction. In Sect.~\ref{dataanalysis}, we report on the spectral and spectro-polarimetric analysis. In Sect.~\ref{discussion} the results are discussed.

\section{Observations and data reduction}\label{datareduction}

{\it IXPE} \citep{weisskopf22} observed NGC~4151 starting on December 8, 2022 with its three Detector Units (DU), for a net exposure time of about 632 ks. The data were calibrated with a standard {\it IXPE} pipeline from the Science Operation Center (SOC).\footnote{\url{https://heasarc.gsfc.nasa.gov/docs/ixpe/analysis/IXPE-SOC-DOC-009-UserGuide-Software.pdf}} The pipeline mainly contains the correction processes on the photoionization events and the track reconstruction process following a standard moments analysis \citep{10.1117/12.459380, fabiani2014astronomical, 2022AJ....163..170D}. In addition, variations on gain properties that are caused by susceptibility of gas status (e.g., temperature and pressure) inside Gas Pixel Detector \citep[GPD,][]{costa01, 2007NIMPA.579..853B, 2012AdSpR..49..143F, 2021APh...13302628B} and non-uniformity of the charging on the Gas Electron Multiplier (GEM) material are accounted for. The onboard calibration data was utilized to deal with the small time scale variations \citep{ferrazzoli2020flight}. The spurious modulation was also taken into account in this process \citep{Rankin_2022}. The scientific analysis was performed using the \textsc{ixpeobssim} software version 30.2.1 \citep{2022SoftX..1901194B}. Source and background data were extracted centered on the source position in the detector frame which covered the entire source emission and source-free regions respectively. For the region selection criteria, we applied a 72\arcsec\ circle for the source and an annulus with an inner and outer radius of 150\arcsec\ and 240\arcsec\ for the background \citep{dimarco23}. In order to estimate the polarization properties, we created (1) the polarization cube (\texttt{PCUBE}) based on the \citet{KISLAT201545} method, which  provided results independent of any spectral modelling, and (2) $I$, $Q$, and $U$ spectra using PHA1, PHAQ1, and PHAU1 algorithm in the \texttt{xpbin} tool inside \textsc{ixpeobssim}. We utilised the version 12 instrument response functions for both methods, which are contained in \textsc{ixpeobssim}. We adopted a minimum of 30 counts binning for spectra $I$ and 0.2 keV constant energy binning for $Q$ and $U$ spectra in order to perform spectro-polarimetric analysis based on $\chi^2$ statistics. For the spectra, we employed the weighted method \citep{2022AJ....163..170D} using the \texttt{alpha075} response matrix to improve the sensitivity of polarimetry measurements. In contrast, this feature is not currently available within the \texttt{pcube} algorithm.

\textit{XMM-Newton} observed NGC~4151 on December 17, 2022 for 50~ks of elapsed time with the EPIC pn \citep{struder01} and the two MOS \citep{turner01} cameras, operating in Small Window and thin filter mode to avoid pile-up effects. Background flares were present during the observation and after the filtering process, the effective exposure time resulted to be of about 33 ks for the pn spectrum. The extraction radii for the source and the background spectra are 20\arcsec\ and 30\arcsec, respectively. The effective area was corrected with the new SAS keyword, \textsc{applyabsfluxcorr}, expressly implemented to provide a better agreement with simultaneous {\it NuSTAR} data. In the following fits, we allow for an energy shift of the order of 1000 km s$^{-1}$ (modelled with \texttt{vashift}) at the iron line to mitigate some residual calibration issues, as noted also in other recent observations of bright AGN (Serafinelli et al., in prep.; Ingram et al., in prep.).

The {\it NuSTAR} \citep{harrison13} observation started on December 16, 2022 simultaneously to {\it XMM} and {\it IXPE} pointings, with both coaligned X-ray telescopes with Focal Plane Module A (FPMA) and B (FPMB). The \texttt{Nupipeline} task and the latest calibration files available in the database (CALDB 20221229) were used to produce and calibrate cleaned event files. In this case the source and background extraction radii are 2\arcmin\ and 1.22\arcmin, respectively. The net exposure time for the FPMA and FPMB resulted to be 97 ks and 96.3 ks. Significant deviations from the pn spectrum were still present in the {\it NuSTAR} spectra below 4 keV, even after applying the correction mentioned above. Therefore, we will consider {\it NuSTAR} data only above 4 keV \citep[see e.g.][]{madsen20}. Moreover, an energy shift between the two instruments is evident at the iron line\footnote{The shift is still present if strictly simultaneous {\it NuSTAR--XMM} spectra are considered, and is relative between the two instruments, so on top of the more modest `absolute' shift described above for the EPIC pn spectrum.}. On the other hand, the MOS data are in agreement with the pn, although with larger uncertainties. In the following fits, we thus applied a linear \textsc{gain fit} of $\sim-60$ eV to the {\it NuSTAR} spectra. We note here that a similar shift is found in AGN observations taken $\sim$1 month before and after our dataset (Serafinelli et al., in prep.; Ingram et al., in prep.).

All the uncertainties are given at 68\% (1$\sigma$) confidence level, unless otherwise stated, while the upper/lower limits are quoted at 99\% (2.6$\sigma$) confidence level for one interesting parameter.
Throughout our analysis, we adopt a redshift $z=0.003326$ for NGC~4151 \citep{2013MNRAS.428.1790W}, and the cosmological parameters $H_0$ = 70 km s$^{-1}$ Mpc$^{-1}$, $\Lambda_0$ = 0.73 and $\Lambda_m$ = 0.27.

\section{Data analysis}\label{dataanalysis}

\subsection{{\it IXPE} polarimetric analysis}\label{pcube}

\begin{figure}
\centering
         \includegraphics[width=1.0\columnwidth]{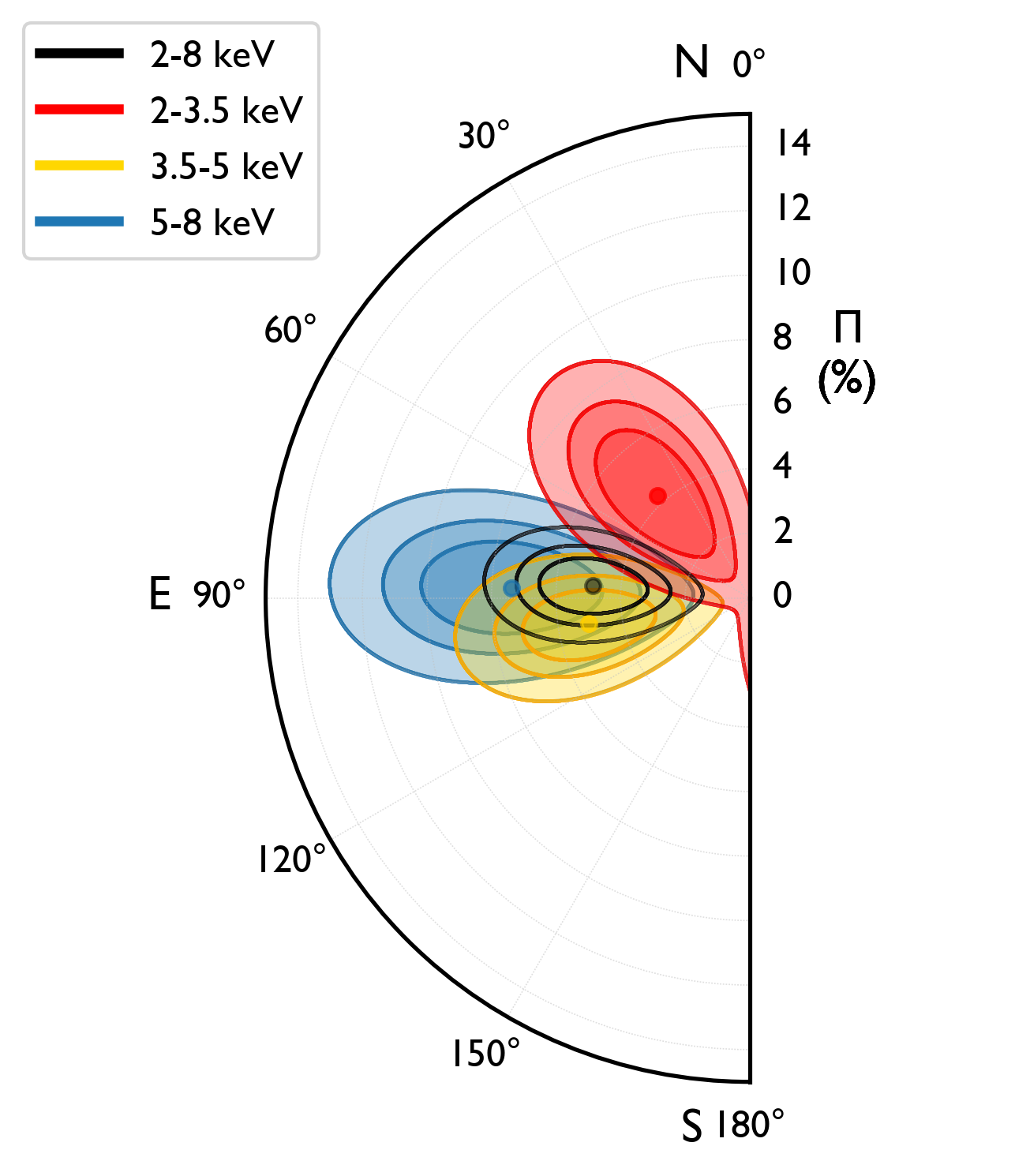}
\caption{Polarization contours (68$\%$, 90$\%$, and 99$\%$ confidence levels for two degrees of freedom) for  the polarization degree $\Pi_{\rm X}$ and the polarization angle $\psi_{\rm X}$ with respect to the north direction. Colors refer to the 2.0--8.0 keV (black), 2.0--3.5 keV (red), 3.5--5.0 keV (yellow), and 5.0--8.0 keV (blue) energy ranges, respectively.} \label{fig:energy_contour}
\end{figure}

We report the first significant polarization detection from NGC~4151 using \texttt{PCUBE} analysis.
The measured polarization parameters from the three combined DUs are $\Pi_{\rm X}= 4.9\% \pm 1.1\%$ and $\psi_{\rm X}= 86\degr\pm 7\degr$ in the 2--8 keV band with background subtraction. The detection significance of these polarization properties is above 99.99$\%$ confidence level ($\sim$ 4.4$\sigma$). In order to examine the energy dependency of the polarization, we tested against the hypothesis that \textit{Q} and \textit{U} Stokes parameters are constant via a $\chi^2$ test, adopting different energy binnings \citep[from 2 to 12 bins over the entire energy band, e.g.][]{2022ApJ...938L...7D}. We found a statistically significant ($>99$\% confidence level) deviation from the constant behaviour in $Q$, when adopting three and four bins. Figure \ref{fig:energy_contour} and Table \ref{tab:pol_energy} show the data into 3 energy bands: 2.0--3.5, 3.5--5.0, and 5.0--8.0 keV. In the polarization contour plot, significant detections are found for the two higher-energy bins (3.5--5.0 and 5.0--8.0 keV), while only a marginal detection can be claimed for the first bin (2.0--3.5 keV), possibly suggesting also a variation of the polarization angle, thus confirming the variability in $Q$ mentioned above. 

\begin{table}
\centering
\caption{Polarization parameters for different energy bands.}\label{t1}
\label{tab:pol_energy}
\begin{tabular}{l c c}
\hline\hline
{Energy range} & {$\Pi_{\rm X} \pm 1\sigma$} & {$\psi_{\rm X} \pm 1\sigma$ } \\
{ (keV)} & {(\%)} & {  (deg)} \\
\hline
2.0 -- 8.0 & 4.9 $\pm$ 1.1 & 86 $\pm$ 7\\[2mm]

2.0 -- 3.5 & 4.3 $\pm$ 1.6 & 42 $\pm$ 11\\
3.5 -- 5.0 & 5.0 $\pm$ 1.4 & 99 $\pm$ 8\\
5.0 -- 8.0 & 7.4 $\pm$ 1.9 & 88 $\pm$ 7\\
\hline
\end{tabular}
\end{table}

\subsection{Spectral Analysis: {\it XMM-Newton} and  {\it NuSTAR}}

The spectral analysis of NGC~4151 is performed with \textsc{xspec} 12.13.0 \citep{Arnaud1996}, taking into account the simultaneous 0.5--10 keV {\it XMM-Newton} and 4--79 keV {\it NuSTAR} spectra. In the new observations, the source has been detected in a high flux state with $F_{\rm{0.5-10 \; keV}}$ $\sim$ 1.7$\times$ 10$^{-10}$ erg s$^{-1}$ cm$^{-2}$.

As mentioned in Sect.~\ref{introduction}, the X-ray spectrum of this source is remarkably complex with different emission and absorption components \citep[e.g.,][]{weaver94,zdziarski96,yang01,derosa07,kraemer08,lubinski10}. Our best fit model implements and tries to simplify those adopted by \citet{keck15} and \citet{szanecki2021} for the previous {\it XMM-Newton}, {\it Suzaku} and {\it NuSTAR} observations of the source. The Galactic absorbing column density, modelled with \texttt{tbabs}, is set to $N_{\rm H}$ = 2.3 $\times$ 10$^{20}$ cm$^{-2}$ \citep{kalberla05} and multiplicative constants (found to be of the order of $\sim1.20$) take into account cross-calibration uncertainties between the two FPM modules and EPIC pn, as well as some flux variability of the source during the longer elapsed time of the {\it NuSTAR} observation. We adopt the default abundance table in \textsc{xspec} \citep{anders89}.

In previous studies, a strong Fe K$\alpha$ emission line with a weak relativistic component was reported in high flux states \citep[e.g.][and references therein]{yaqoob95,zoghbi19}. In the new {\it XMM-Newton} and {\it NuSTAR} observations, the Fe K$\alpha$ line profile is well modelled by a single Gaussian with an equivalent width of EW = $100 \pm 6$ eV and a resolved width of $\sigma=40\pm10$ eV. Such a width is routinely measured in high-quality spectra both of obscured and unobscured AGN, suggesting an origin of the line in the broad line region or in the torus \citep[e.g.][]{Shu2010,Shu2011}.
Therefore in the following, also to simplify the spectro-polarimetric fit presented in Sect.~\ref{spectropolarimetry}, we will not include any relativistic reflection component in our model, as instead used in \citet{keck15} and \citet{szanecki2021}. In any case, we note here that we verified that including a relativistic reflection component to our broadband model gives a similar fit, not affecting in a significant way the other main parameters of the model\footnote{An equivalent fit ($\Delta\chi^2=+2$ for the same d.o.f.) is obtained by instead deconvolving the reflection components with \texttt{rdblur}, which introduces relativistic effects from an accretion disc around a non-rotating black hole \citep{fabian89}. As expected, given the modest broadening of the line, the best fit inner radius is very large ($r_{\rm in}=110^{+40}_{-20}$ $r_\mathrm{g}$, where $r_\mathrm{g} = GM/{c^2}$ is the gravitational radius), and the inclination very low ($i=3^{+5}_{-2}$~deg). 
All the other parameters are the same, within errors, with respect to those of the best fit.}.

We model the primary continuum with a thermally Comptonized continuum \citep[\texttt{nthcomp} in \textsc{xspec},][]{zdziarski96, zycki99}, assuming seed photons from a disc-blackbody with temperature fixed at $kT_{\rm bb} =8$ eV \citep[as expected from a standard accretion disc with the BH mass and observed luminosity of NGC~4151,][]{shakura73}.
For the reflection component, we used \textsc{\texttt{BORUS}} \citep{balokovic18,balokovic19}, which models reprocessing from a torus with variable covering factor, self-consistently illuminated by a \texttt{nthcomp} spectrum with the same photon index and electron temperature as the primary continuum. In view of the joint spectro-polarimetric fit with the {\it IXPE} data presented in Sect.~\ref{spectropolarimetry}, we separated the reflection component from the corresponding fluorescent lines with the two dedicated \texttt{BORUS} tables, linking all the parameters (including the normalization). Since they are not constrained by the fit, we fixed the cosine of the inclination and the covering factor of the torus to 0.6 (appropriate for an intermediate Seyfert galaxy) and 0.5 (default value in \texttt{BORUS}), respectively. 
Moreover, in order to account for the modest broadening of the iron line as reported above, we convolve the \texttt{BORUS}  tables with a \texttt{gsmooth} of $28^{+15}_{-17}$ eV.

The X-ray spectrum of NGC~4151 is well-known to be strongly affected by complex absorption. Similarly to \citet{keck15}, we used two partial-covering neutral absorbing layers (PC, modelled by \texttt{zpcfabs}) and a warm absorber (WA, modelled with \texttt{zxipcf} and covering factor fixed at 1). The presence of a second WA phase at lower ionization, which has been observed in some prior studies \citep[e.g.][]{keck15,zoghbi19,szanecki2021}, is not required by the data ($\Delta\chi^2=-3$ for 2 d.o.f. less). However, it is plausible that one of the neutral PCs may already mimic a low-ionization WA.
Finally, as suggested by the presence of emission lines at $\sim$0.5 and $\sim$ 0.9 keV and by previous results based on high-resolution spectra \citep[][]{Schurch2004,Guainazzi2007,Bianchi2019}, the remaining soft X-ray emission is modelled with a photoionized plasma emission component, produced with CLOUDY \citep{ferland98} closely to what described in \cite{Bianchi2010a}. Some further residuals due to an imperfect modelization of the photoionized gas around the \ion{O}{vii} emission line triplet at $\sim0.5$ keV are modelled with a Gaussian component.

In summary, our model can be written in \textsc{xspec} as \code{(tbabs)*(CLOUDY + zgauss + PC*PC*WA*(gsmooth*({BORUS} 1 + {BORUS} 2) + nthcomp))}. This gives a good representation of the {\it XMM-Newton} + {\it NuSTAR} data with $\chi^2$/d.o.f = 743/660.

\begin{figure*}
\centering
    \includegraphics[width=0.47\textwidth]{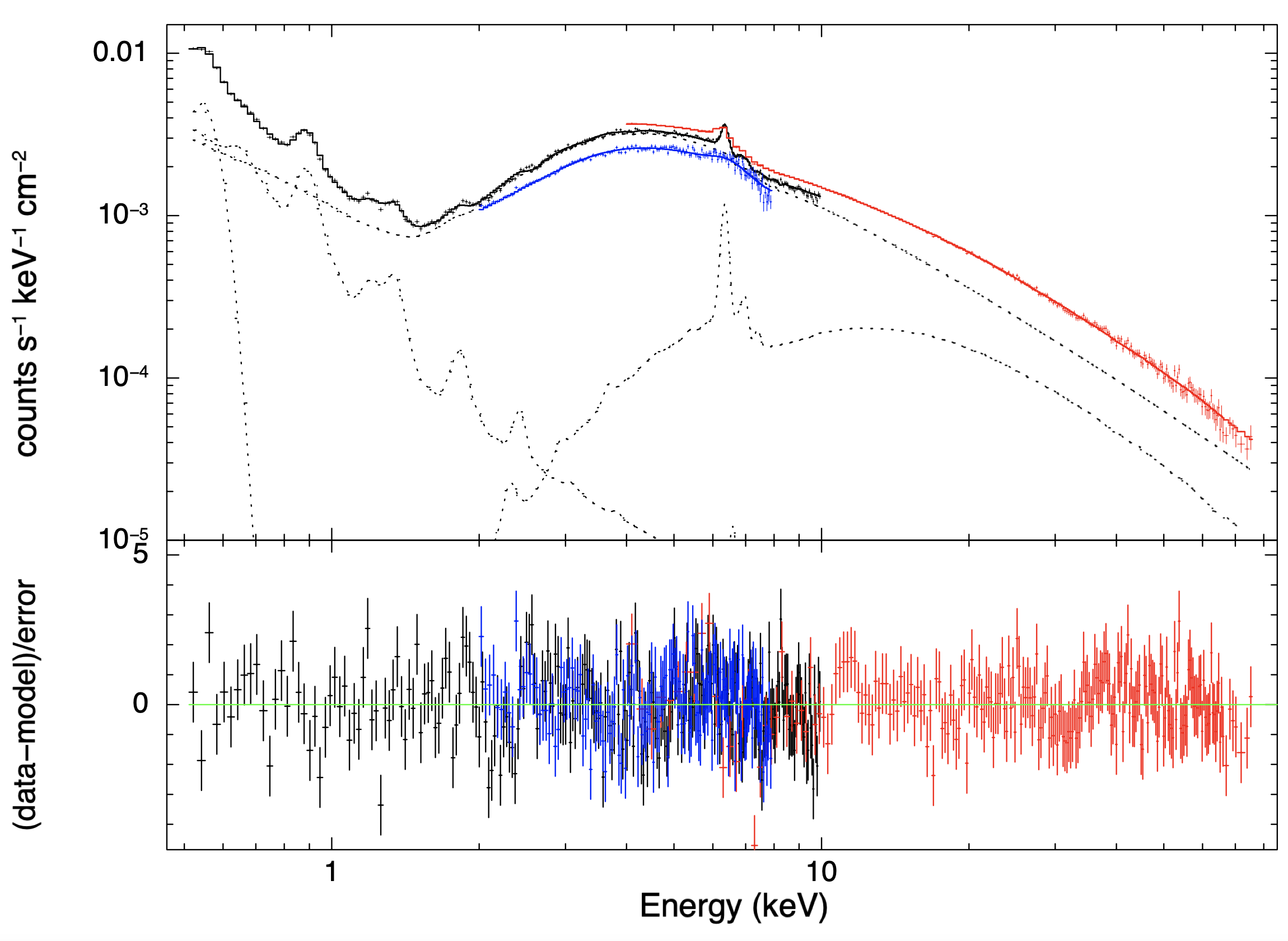}
    \includegraphics[width=0.49\textwidth]{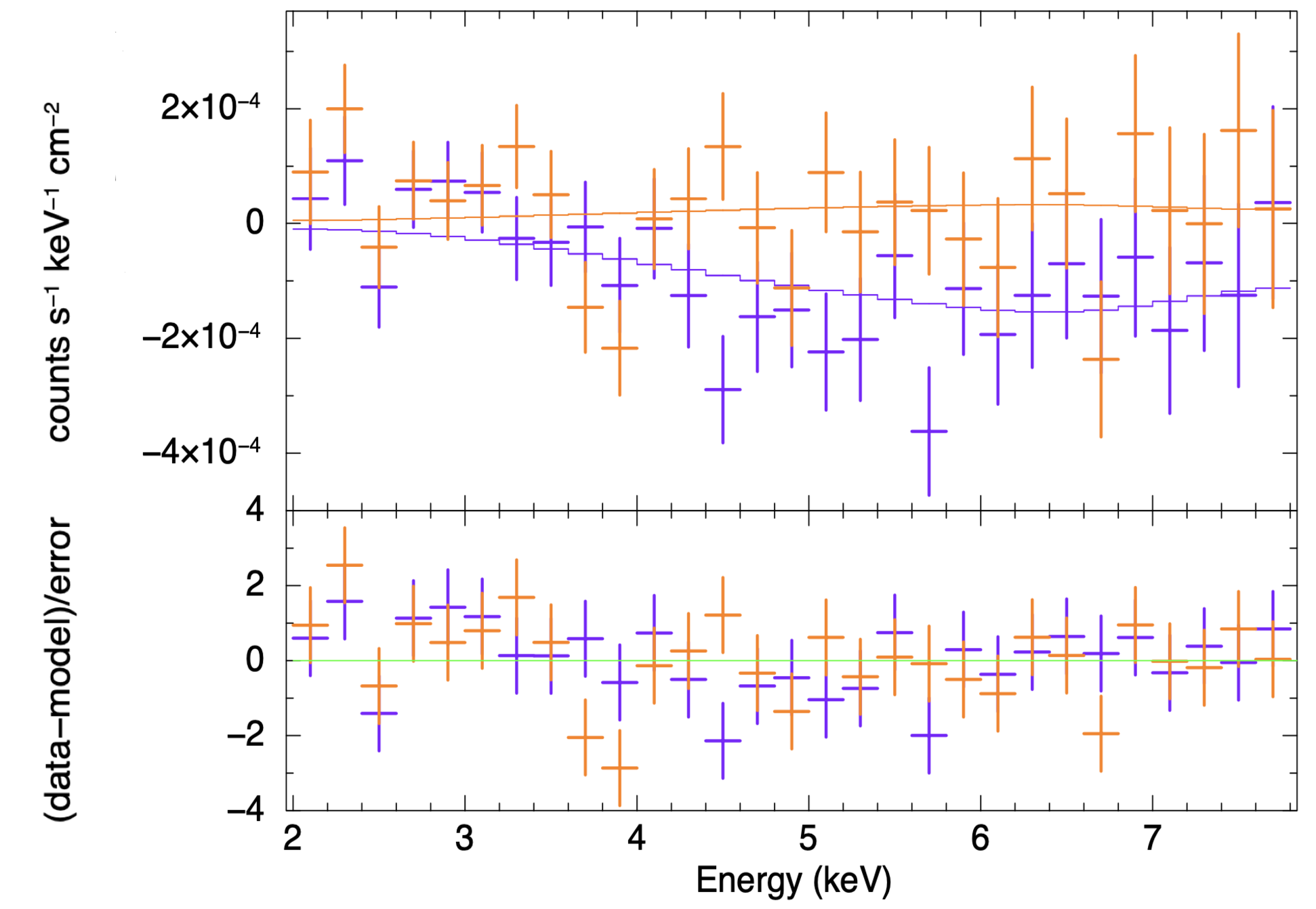}
  \caption{{\it Left panel:} \textit{XMM-Newton}/EPIC pn (in black), grouped {\it NuSTAR} FPMA and FPMB (in red), {\it IXPE} grouped Stokes $I$  (in blue) simultaneous spectra of NGC~4151 with residuals. The dashed lines represent the different model components.  {\it Right-panel:} $Q$ (in purple) and $U$ (in orange) grouped Stokes spectra are shown with residuals.}
  \label{fig:1}
\end{figure*}

 The best fit parameters are reported in Table \ref{tab:1}, while spectra and residuals are shown in Fig.~\ref{fig:1}. It is important to note that in this model the contribution of the reflection component to the total 2--8 keV flux is of the order of 6 per cent, reaching up to $\sim$ 16 per cent in the 6--8 keV band, due to the presence of the Fe K$\alpha$ line. In the 2--3.5 keV band, no contribution from the photoionized emission is present, since it becomes significant only at lower energies (see Fig.~\ref{fig:1}). On the other hand, in this band there is an excess with respect to the absorbed primary continuum, which our model treats as the leakage of the primary emission through the partial coverers. Any other component, used to model this excess, would contribute up to 20 per cent of the total flux in the 2--3.5 keV band. 

\subsection{\label{spectropolarimetry}Spectro-polarimetric analysis: {\it XMM-Newton}, {\it NuSTAR} and {\it IXPE}}

We added the {\it IXPE} data ($I$, $Q$ and $U$ spectra of the three detectors) to the {\it XMM-Newton}+{\it NuSTAR} best fit presented above, with all the spectral parameters linked to the other instruments, allowing only an inter-calibration constant (found to be of the order of $\sim0.80$) for each detector to vary. We then add separate \code{polconst} multiplicative models to account for the polarization of each additive component of the global model. The polarization degree $\Pi$ and angle $\Psi$ are set to 0 for the \texttt{BORUS} component producing the emission lines, since they are expected to be intrinsically not polarized, as well as for the CLOUDY component, which does not contribute at all in the {\it IXPE} energy band (see previous Section). On the other hand, the primary Comptonized continuum and the reflection component associated to the other \texttt{BORUS} table have $\Pi$ and $\Psi$ free to vary. 
The best fit gives $\chi ^2$/d.o.f= 1433/1264, with no appreciable variations in the spectral parameters with respect to the one without \textit{IXPE}, being indeed dominated by the much higher sensitivity, spectral resolution and broad band coverage of {\it XMM-Newton} and {\it NuSTAR}. 
Considering the complex X-ray spectrum of NGC~4151 with multiple components in absorption and emission, the cross-calibration uncertainties among the several different instruments used, the variability of the source and the much longer exposure time of {\it IXPE} with respect to {\it XMM-Newton} and {\it NuSTAR}, we find this fit, if not ideal, acceptable given the goal of the paper. We note that the fit applied only to the \textit{IXPE} data gives a significantly better fit ($\chi ^2$/d.o.f= 672/612), confirming that the model is a good representation of the data in the 2-8 keV band, where all the polarimetric information is present. Moreover, the polarimetric parameters themselves are only sensitive to the IXPE data, and are not directly affected by the spectroscopic fit on the other data-sets.

In this configuration, we get loose constraints for the polarimetric parameters, i.e. $\Pi<5\%$  and unconstrained angle for the primary continuum, and $\Pi>38\%$ and $\Psi=96\degr\pm16\degr$ for the reflection component. We thus linked the angles of the two components, either to be equal or to differ by 90\degr. In both cases, we still obtain that the polarization is dominated by the reflection component, and the polarization degree of the primary emission is an upper limit. We therefore fixed the polarization properties of the reflection component to physically motivated values, 15, 20, 30 per cent \citep[as found, for example, for the reflection-dominated Circinus galaxy,][]{ursini23}, constraining its polarization angle to be at 90\degr\ with respect to that of the primary emission. The resulting fits are marginally worse than the best fit, the $\chi^2$ being 1452, 1453 and 1455 for 1266 d.o.f., respectively, and the polarization degree of the primary continuum is now constrained at $\Pi=4.1\pm0.8$\%, $\Pi=4.3\pm0.8$\% and $\Pi=4.6\pm0.8$\% with polarization angles $\Psi=82\degr\pm7\degr$, $\Psi=81\degr\pm7\degr$ and $\Psi=80\degr\pm8\degr$, respectively.

These results may be driven by the lower $\Pi$ and different $\Psi$ observed in the 2--3.5 keV band (see Table~\ref{tab:pol_energy}), possibly due to another spectral component which dilutes the polarization of the primary continuum. We therefore modified the best fit model with a spectroscopically equivalent one, but decoupling the soft X-ray emission leaking through the partial coverers from the primary emission, which allows us to assign another \texttt{polconst} to this component.\footnote{Each \code{zpcfabs} has been replaced by the equivalent expression \code{$c$*zphabs +(1-$c$)}, where $c$ is the same covering factor determined in the best fit.}  We set its $\Pi$ and $\Psi$ to 0 -- this is physically motivated by the possibility that the leaked continuum comes from a variety of line-of-sights, or that this further soft component has instead another origin, independent from the primary continuum and not taken into account by our modelling. 
Interestingly, the observed variability of NGC~4151 peaks at 2-3 keV, where it is indeed significantly larger than at higher energies \citep[e.g.][]{Igo2020,Beuchert2017}. This can be attributed to strong absorption variability, but may further suggest the presence of another spectral component.
In this new configuration, with the two polarization angles forced to differ by 90\degr, the best fit is statistically equivalent to the initial one ($\chi ^2$/ d.o.f = 1441/1265), but now all the polarization is attributed to the primary continuum, with $\Pi=7.7\pm1.5$\% and $\Psi=87\degr\pm6\degr$, while only an upper limit is found for the reflection component $\Pi<27\%$. A similar result is obtained if the polarization angles are forced to be the same, but the polarization degree of the reflection component is completely unconstrained in this case.

\begin{table}
\caption{Best fit parameters from the spectro-polarimetric analysis.}
\label{tab:1}
\renewcommand{\arraystretch}{1.1}
\begin{tabular}{lllll}
\hline\hline
{Parameter} & {Value} \\
\hline
\multicolumn{2}{c}{\texttt{CLOUDY \footnotesize{(Photoionized emitter)}}}\\
$\log U$ & 1.35 $\pm$ 0.01 \\
$\log (N_\mathrm{H}$ / cm$^{-2}$) & 21.63 $\pm$ 0.02 \\
\multicolumn{2}{c}{\texttt{PC 1 \footnotesize{(Neutral absorber 1)}}} \\
$\log (N_\mathrm{H}$ /cm$^{-2}$) & 23.02 $\pm$ 0.01 \\ 
C$\mathrm{f}$ & 0.78 $\pm$ 0.01 \\
\multicolumn{2}{c}{\texttt{PC 2 \footnotesize{(Neutral absorber 2)}}} \\
$\log (N_\mathrm{H}$ / cm$^{-2}$) & 22.64 $\pm$ 0.01 \\
C$\mathrm{f}$ & 0.95 $\pm$ 0.01 \\
\multicolumn{2}{c}{\texttt{WA \footnotesize{(Warm absorber)}}} \\
$\log(N_\mathrm{H}$ / cm$^{-2}$) & 23.13 $\pm$ 0.03 \\
$\log (\xi$ / erg cm s$^{-1}$) & 4.12 $\pm$ 0.02 \\
\multicolumn{2}{c}{\texttt{\texttt{BORUS}  1/2 \footnotesize{(Neutral reflector 1/2)}}} \\
$\log (N_\mathrm{H}$ / cm$^{-2}$) & $24.45 \pm 0.01$ \\
$A_{\mathrm{Fe}}$ & 0.62 $\pm$ 0.01 \\
norm & 0.09 $\pm$ 0.01 \\
\multicolumn{2}{c}{\texttt{nthcomp \footnotesize{(Comptonized primary continuum)}}} \\
$\Gamma$  & 1.85 $\pm$ 0.01  \\
$kT_{\mathrm{e}}$ [keV] & $60^{+7}_{-6}$ \\
norm & 0.09 $\pm$ 0.01  \\
\\
$\log (F_{\mathrm{2-10 \; keV}}$ / erg cm$^{-2}$ s$^{-1}$) & $-$9.78 $\pm$ 0.01 \\
$\log (L_{\mathrm{2-10 \; keV}}$ / erg s$^{-1}$) & 42.61 $\pm$ 0.01 \\
\hline
\end{tabular}
\end{table}

\section{Discussion}\label{discussion}

\subsection{The geometry of the X-ray corona}

The high measured polarization degree of NGC~4151, determined by both the model-independent and spectro-polarimetric analysis, is driven by the signal at higher energies, where the emission from the hot corona dominates. This immediately excludes a `spherical' lamppost along the disc axis as a possible geometry for the hot corona in this source. Indeed, such a geometry is very symmetric, so the polarization degree is expected to be lower than $1-3$\%, even for very high inclinations \citep[e.g.][]{poutanen96,tamborra18,ursini22}. Moreover, the corresponding polarization angle is expected to be perpendicular to the disc axis, while the measured $\Psi$ is in the direction of the radio emission \citep[$\sim$ 83\degr,][and references therein]{Harrison1986,ulvestad98}, suggesting instead that the polarization occurs on the equatorial plane. Two other possible coronal geometries are viable and will be considered here: a \textit{slab} extending above and below the accretion disc, and a \textit{wedge}, in which the accretion disc is truncated and the X-ray corona acts as hot accretion flow extending down to the innermost stable circular orbit (ISCO) with a defined opening angle (Tagliacozzo et al., in prep.). The \textit{slab} geometry is investigated even if it is known to produce relatively soft spectra ($\Gamma \geq$ 2) when radiative equilibrium between the disc and the corona is established (e.g. \citealt{haardt93,stern95} and relevant discussion in \citealt{poutanen2018}). The somewhat harder photon index observed in NGC~4151 can still be accommodated with this geometry assuming that the cold accretion disc is truncated at some radius and the inner part is occupied by the hot accretion flow. 
The seed photons for Comptonization in this case may come from the outer cold disc or be internal synchrotron photons \citep{Veledina11}. It is very difficult to distinguish between these two scenarios, because in both cases photons undergo many scatterings before they reach the \textit{IXPE} energy band.
 
We followed the approach of \citet{ursini22}, performing various simulations with the two geometries, using the general relativistic Monte Carlo radiative transfer code \code{MONK} \citep{zhang19}. We have also cross-checked these results with those obtained with an iterative radiation transport solver \citep{poutanen96,veledina22}.
In our simulations, we assumed a BH mass of $4.57 \times 10^7 M_\odot$ \citep{bentz06}, a spin $a = 0.998$ and an Eddington ratio $L_{\rm{Bol}}/L_{\rm{Edd}}$ = 1\%. As for the corona, we adopted a temperature of 60 keV, as derived from our spectral analysis (see Table~\ref{tab:1}), and the Thomson optical depth (defined with respect to the half-thickness of the slab/wedge\footnote{This definition is the same as in \code{compTT} \citep{titarchuk94}, while it is half that of \code{compPS} in the standard configuration \citep{poutanen96}.}) $\tau=0.5$, which reproduces the observed photon index ($\Gamma=1.85$) in both geometries for the given temperature.
For both geometries, we consider the X-ray corona inner radius at the ISCO ($R_{\mathrm{in}}$ = 1.24 $r_\mathrm{g}$), while the outer radius is at $100$ $r_\mathrm{g}$ for the \textit{slab} geometry, and coincides with the inner radius of the accretion disc, $R_{\mathrm{out}}$ = $R_{\mathrm{in}}^{\mathrm{disc}}$ = 25 $r_\mathrm{g}$, for the \textit{wedge}. For the latter geometry, the tested opening angles are 30\degr, 45\degr\ and 60\degr. The height of the slab is 1 $r_\mathrm{g}$.

\begin{figure}
    \includegraphics[width=0.5\textwidth]{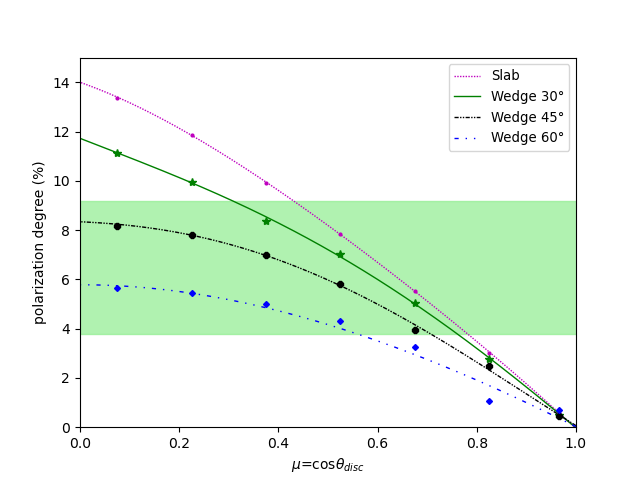}
  \caption{Monte Carlo simulations performed with the Comptonization code {\sc monk}. Both \textit{wedge} and \textit{slab} geometry have been considered. The light green band shows the polarization degree range resulting from the model-independent (see Sect.~\ref{pcube}) and the spectro-polarimetric analysis (see Sect.~\ref{spectropolarimetry}). We adopt for all the simulations $kT_{\rm e} = 60$~keV and $\tau=0.5$. The obtained PA is always parallel to the disc axis. See text for details.}
  \label{fig:2}
\end{figure}

It is worth stressing that in all cases, the expected polarization angle is parallel to the disc axis, so in agreement with the observed one. 
The resulting polarization degree is shown in Figure \ref{fig:2}, as a function of the  cosine of the inclination angle with respect to the observer. It is clear that the observed polarization degree in NGC~4151 is well reproduced in all cases, only assuming moderate inclinations ($i\gtrsim 40\degr-50\degr$), as reasonable for an intermediate Seyfert galaxy.
In particular, the inclination results to be more constrained in the case of the \textit{slab} and for low opening angles of the \textit{wedge}, being in the range $40\degr-70\degr$. On the other hand, for larger opening angles of the \textit{wedge}, the required inclination can be higher. Note that these different geometries also agree with the lack of a significant variation of the polarization degree with energy \citep[e.g.,][]{ursini22} in agreement with the observations (see Sect. \ref{pcube} and Table \ref{tab:pol_energy}).

The disc inclination in NGC~4151 is very uncertain \citep{marin16}. It ranges from $i = 0\degr$ to 33\degr\  when estimated via the relativistic reflection component in the X-rays \citep{nandra97,Keck2015,Beuchert2017,Miller2018}, but a much more inclined system ($\sim$58\degr) is suggested by reverberation studies of the broad-line region (BLR) \citep{bentz22}. The mismatch between the various values comes from both the technique and the location of the probed region (the disc or the BLR). \citet{Miller2018} suggested that a warp between the innermost and outer part of the accretion disc in NGC~4151 might resolve this apparent discrepancy in inclination. However, it does not fit the \textit{IXPE} results. In fact, the inclination estimated from BLR reverberation studies  matches better the one obtained from the X-ray polarization. A more systematic analysis of bright and nearby Seyfert-1s is needed to verify this conclusion.

\subsection{Comparison to lower energies polarization}

\citet{Marin2020} presented the most extensive review of the ultraviolet, optical and infrared linear continuum polarization of NGC~4151. From the ultraviolet to the near-infrared ($\le$ 1~$\mu$m), the polarization degree is wavelength-dependent but does not exceed 2\%, while the polarization position angle remains constant at 80\degr--90\degr. Because the polarization angle is parallel to the parsec-scale radio axis, NGC~4151 optical polarization emerges from reprocessing along the equatorial plane. \citet{Gaskell2012} proved, using polarization reverberation mapping, that the polarization emerges from scattering in a flattened region within the low-ionization component of the BLR. The time-lag and polarization angle are inconsistent with both scattering onto the dusty torus and with an intrinsic polarization of the continuum. However, the polarized light spectrum of NGC~4151 appears to corroborate the existence of an optically thick, thermally heated accretion disc structure, at least in its outer near-IR emitting radii \citep{Marin2020}. In the infrared, a smooth rotation of the polarization position angle down to $\sim$ 45\degr\ indicates the onset of dichroic absorption from aligned dust grains in the torus. We note that a similar polarization angle is found here in the 2--3.5 keV band ($\psi_{\rm X}=42\pm11$\degr, see Table~\ref{t1}), although with lower statistical confidence than at higher energies. Interestingly, this angle would be in agreement with the extended narrow-line region observed with {\it Chandra} \citep{wang11} and {\it HST} ($45\degr\pm5\degr$, \citealt{evans93,Das2005}). 

The X-ray polarization degree we measured is different from the archival and contemporaneous\footnote{\textit{B}, \textit{V}, \textit{R} and \textit{I} broadband polarimetry of NGC~4151 was obtained contemporaneously to the \textit{IXPE} observation using the Perkins Telescope observatory. Those new measurements are in all respects similar to the archival data and will be presented in a future publication compiling recent optical polarimetric data from Changing Look AGNs.} ultraviolet, optical and infrared polarization of NGC~4151. It indicates that X-ray polarization comes from a different region than the BLR or the torus, and is indeed consistent with an origin in a slab-like corona. The polarization angle, however, is the same as in the ultraviolet and optical, indicating that reprocessing mainly occurs along the equatorial plane from the X-rays to the near-infrared. A deeper analysis of the X-ray to infrared polarization of NGC~4151 will be presented in a future paper. Future \textit{IXPE} observations catching the source at different flux and spectral states will be crucial to further disentangle the contribution of each spectral component to the observed X-ray polarization.

\section*{Acknowledgements}

The Imaging X-ray Polarimetry Explorer ({\it IXPE}) is a joint US and Italian mission. The US contribution is supported by the National Aeronautics and Space Administration (NASA) and led and managed by its Marshall Space Flight Center (MSFC), with industry partner Ball Aerospace (contract NNM15AA18C). The Italian contribution is supported by the Italian Space Agency (Agenzia Spaziale Italiana, ASI) through contract ASI-OHBI-2017-12-I.0, agreements ASI-INAF-2017-12-H0 and ASI-INFN-2017.13-H0, and its Space Science Data Center (SSDC) with agreements ASI- INAF-2022-14-HH.0 and ASI-INFN 2021-43-HH.0, and by the Istituto Nazionale di Astrofisica (INAF) and the Istituto Nazionale di Fisica Nucleare (INFN) in Italy. This research used data products provided by the {\it IXPE} Team (MSFC, SSDC, INAF, and INFN) and distributed with additional software tools by the High-Energy Astrophysics Science Archive Research Center (HEASARC), at NASA Goddard Space Flight Center (GSFC). We thank the {\it XMM-Newton} and {\it NuSTAR} SOCs for granting and performing the respective observations of the source. Part of the French contribution is supported by the Scientific Research National Center (CNRS) and the French Space Agency (CNES).

\section*{Data Availability}

The {\it IXPE} data used in this paper are publicly available in the HEASARC database (\url{https://heasarc.gsfc.nasa.gov/docs/ixpe/archive/}). The {\it XMM-Newton} and {\it NuSTAR} data underlying this article are subject to an embargo of 12 months from the date of the observations. Once the embargo expires the data will be publicly  available from the {\it XMM-Newton} science archive (\url{http://nxsa.esac.esa.int/}) and the {\it NuSTAR} archive (\url{https://heasarc.gsfc.nasa.gov/docs/ nustar/nustar_archive.html}). The MONK simulation data supporting the findings of the article will be shared on reasonable request.


\bibliographystyle{mnras}
\bibliography{biblio} 


\vspace{1cm}

\noindent \textbf{Affiliations:} \\
\noindent
\textit{
     $^{1}$Université Grenoble Alpes, CNRS, IPAG, 38000 Grenoble, France \\
     $^{2}$Dipartimento di Matematica e Fisica, Università degli Studi Roma Tre, Via della Vasca Navale 84, 00146 Roma, Italy \\
     $^{3}$INAF Istituto di Astrofisica e Planetologia Spaziali, Via del Fosso del Cavaliere 100, I-00133 Roma, Italy\\
     $^{4}$Dipartimento di Fisica, Universit\`a degli Studi di Roma “La Sapienza”, Piazzale Aldo Moro 5, I-00185 Roma, Italy\\
     $^{5}$Dipartimento di Fisica, Università degli Studi di Roma “Tor Vergata”, Via della Ricerca    Scientifica 1, I-00133 Roma, Italy\\
     $^{6}$Instituto de Astrofísica de Andalucía—CSIC, Glorieta de la Astronomía s/n, 18008 Granada, Spain \\
     $^{7}$Department of Physics and Kavli Institute for Particle Astrophysics and Cosmology, Stanford University, Stanford, California 94305, USA \\
     $^{8}$Université de Strasbourg, CNRS, Observatoire Astronomique de Strasbourg, UMR 7550, 67000 Strasbourg, France \\
     $^{9}$ASI - Agenzia Spaziale Italiana, Via del Politecnico snc, 00133 Roma, Italy \\
     $^{10}$Space Science Data Center, Agenzia Spaziale Italiana, Via del Politecnico snc, 00133 Roma, Italy \\
     $^{11}$Istituto Nazionale di Fisica Nucleare, Sezione di Roma "Tor Vergata", Via della Ricerca Scientifica 1, 00133 Roma, Italy \\
     $^{12}$Department of Astronomy, University of Maryland, College Park, Maryland 20742, USA \\
     $^{13}$School of Mathematics, Statistics, and Physics, Newcastle University, Newcastle upon Tyne NE1 7RU, UK \\
     $^{14}$Department of Physics and Astronomy, 20014 University of Turku, Finland \\
     $^{15}$MIT Kavli Institute for Astrophysics and Space Research, Massachusetts Institute of Technology, 77 Massachusetts Avenue, Cambridge, MA 02139, USA \\
     $^{16}$Astronomical Institute of the Czech Academy of Sciences, Bo{\v c}ní II 1401/1, 14100 Praha 4, Czech Republic \\
     $^{17}$Astronomical Institute, Charles University, V Hole{\v s}ovi{\v c}kách 2, CZ-18000 Prague, Czech Republic \\
     $^{18}$Nordita, KTH Royal Institute of Technology and Stockholm University, Hannes Alfvéns väg 12, SE-10691 Stockholm, Sweden \\
     $^{19}$National Astronomical Observatories, Chinese Academy of Sciences, 20A Datun Road, Beijing 100101, China \\
     $^{20}$INAF Osservatorio Astronomico di Roma, Via Frascati 33, 00078 Monte Porzio Catone (RM), Italy \\
     $^{21}$INAF Osservatorio Astronomico di Cagliari, Via della Scienza 5, 09047 Selargius (CA), Italy \\
     $^{22}$Istituto Nazionale di Fisica Nucleare, Sezione di Pisa, Largo B. Pontecorvo 3, 56127 Pisa, Italy \\
     $^{23}$Dipartimento di Fisica, Università di Pisa, Largo B. Pontecorvo 3, 56127 Pisa, Italy \\
     $^{24}$NASA Marshall Space Flight Center, Huntsville, AL 35812, USA \\
     $^{25}$Istituto Nazionale di Fisica Nucleare, Sezione di Torino, Via Pietro Giuria 1, 10125 Torino, Italy \\
     $^{26}$Dipartimento di Fisica, Università degli Studi di Torino, Via Pietro Giuria 1, 10125 Torino, Italy \\
     $^{27}$INAF Osservatorio Astrofisico di Arcetri, Largo Enrico Fermi 5, 50125 Firenze, Italy \\
     $^{28}$Dipartimento di Fisica e Astronomia, Università degli Studi di Firenze, Via Sansone 1, 50019 Sesto Fiorentino (FI), Italy \\
     $^{29}$Istituto Nazionale di Fisica Nucleare, Sezione di Firenze, Via Sansone 1, 50019 Sesto Fiorentino (FI), Italy \\
     $^{30}$Science and Technology Institute, Universities Space Research Association, Huntsville, AL 35805, USA \\
     $^{31}$Institut für Astronomie und Astrophysik, Universität Tübingen, Sand 1, 72076 Tübingen, Germany \\
     $^{32}$ RIKEN Cluster for Pioneering Research, 2-1 Hirosawa, Wako, Saitama 351-0198, Japan\\
     $^{33}$California Institute of Technology, Pasadena, CA 91125, USA\\
     $^{34}$Yamagata University, 1-4-12 Kojirakawamachi, Yamagatashi 990-8560, Japan \\
     $^{35}$University of British Columbia, Vancouver, BC V6T 1Z4, Canada \\
     $^{36}$International Center for Hadron Astrophysics, Chiba University, Chiba 263-8522, Japan \\
     $^{37}$Institute for Astrophysical Research, Boston University, 725 Commonwealth Avenue, Boston, MA 02215, USA \\
     $^{38}$Department of Astrophysics, St. Petersburg State University, Uni- versitetsky pr. 28, Petrodvoretz, 198504 St. Petersburg, Russia \\
     $^{39}$Department of Physics and Astronomy and Space Science Center, University of New Hampshire, Durham, NH 03824, USA \\
     $^{40}$Physics Department and McDonnell Center for the Space Sciences, Washington University in St. Louis, St. Louis, MO 63130, USA \\
     $^{41}$Finnish Centre for Astronomy with ESO, 20014 University of Turku, Finland \\
     $^{42}$Graduate School of Science, Division of Particle and Astrophysical Science, Nagoya University, Furocho, Chikusaku, Nagoya, Aichi 464-8602, Japan \\
     $^{43}$Hiroshima Astrophysical Science Center, Hiroshima University, 1- 3-1 Kagamiyama, Higashi-Hiroshima, Hiroshima 739-8526, Japan \\
     $^{44}$University of Maryland, Baltimore County, Baltimore, MD 21250, USA \\
     $^{45}$NASA Goddard Space Flight Center, Greenbelt, MD 20771, USA \\
     $^{46}$Center for Research and Exploration in Space Science and Technology, NASA/GSFC, Greenbelt, MD 20771, USA \\
     $^{47}$Department of Physics, The University of Hong Kong, Pokfulam, Hong Kong \\ 
     $^{48}$Department of Astronomy and Astrophysics, Pennsylvania State University, University Park, PA 16802, USA \\
     $^{49}$Center for Astrophysics | Harvard \& Smithsonian, 60 Garden St, Cambridge, MA 02138, USA \\
     $^{50}$INAF Osservatorio Astronomico di Brera, Via E. Bianchi 46, 23807 Merate (LC), Italy \\
     $^{51}$Dipartimento di Fisica e Astronomia, Università degli Studi di Padova, Via Marzolo 8, 35131 Padova, Italy \\
     $^{52}$Mullard Space Science Laboratory, University College London, Holmbury St Mary, Dorking, Surrey RH5 6NT, UK \\
     $^{53}$Anton Pannekoek Institute for Astronomy \& GRAPPA, University of Amsterdam, Science Park 904, 1098 XH Amsterdam, The Netherlands \\
     $^{54}$Guangxi Key Laboratory for Relativistic Astrophysics, School of Physical Science and Technology, Guangxi University, Nanning 530004, China \\
}

\bsp	
\label{lastpage}
\end{document}